\documentclass[aps, pra, a4paper, amsfonts, amssymb, amsmath, 
reprint, showkeys, nofootinbib, twoside, longbibliography]{revtex4-1}
\usepackage{graphicx} 
\usepackage[english]{babel}
\usepackage[utf8]{inputenc}
\usepackage[colorinlistoftodos, color=green!40, prependcaption]{todonotes}
\usepackage{amsthm}
\usepackage{mathtools}
\usepackage{physics}
\usepackage{xcolor}
\usepackage{graphicx}
\usepackage[left=23mm,right=13mm,top=35mm,columnsep=15pt]{geometry} 
\usepackage{adjustbox}
\usepackage{placeins}
\usepackage[T1]{fontenc}
\usepackage{lipsum}
\usepackage{csquotes}
\usepackage{soul}
\usepackage[pdftex, pdftitle={Article}, pdfauthor={Author}]{hyperref} 
\usepackage{enumitem}

\begin{document}

\title{Ontic and epistemic states in the theory of spacetime-local beables}
\author{Nathan Argaman}
    \email[Correspondence email address: ]{argamann@post.bgu.ac.il}
    \affiliation{Department of Physics, Ben-Gurion University of the Negev, Be'er Sheva, Israel {\rm and} \\Department of Physics, Nuclear Research Center -- Negev, P.O. Box 9001, Be'er Sheva 84190, Israel}

\begin{abstract}
{Bell's theorem rules out developing  a locally causal theory to describe quantum phenomena.  Many take this to imply that any model of quantum entanglement must employ variables (called beables by Bell) which follow nonlocal rules, even though signaling is local.  The~alternative is to adopt an all-at-once (block universe) approach, with beables which may depend on both past and future inputs, even though signaling is causal.
Within this lenient-causality approach (a.k.a.~retrocausal), simple cases of entanglement have been successfully described by locally mediated stochastic toy models{, i.e.,~toy models which are local in a sense which generalizes Bell's local causality.  Developing a widely applicable reformulation of quantum mechanics along these lines is a grand challenge.  This work presents a general framework for such models and theories, and identifies the corresponding ontic and epistemic states.  The epistemic state is closely}  analogous to the quantum state, yielding an explanation for the collapse of the wavefunction.  In the case of the models of the framework, it is clear what the information is about.  {The expression for the empirically verifiable predictions of the models in terms of the ontic and epistemic states displays remarkable} parallels to the Born rule.
A toy-model example is discussed.}
\end{abstract}

\maketitle

\section{Introduction}

A century ago, Born's rule for the probabilities of quantum events heralded a departure from determinism in physics.  As is well known, this was not accepted by Einstein. A~decade later, Einstein, Podolsky and Rosen (EPR) provided a prima facie proof that quantum mechanics (QM) was incomplete, and that determinism would be restored if one could find a deeper-level, hidden-variable description of the properties of the particles for which QM gives only probabilities~\cite{EPR1935}.  This EPR proof was based on accepting the predictions of QM, and on local causality: for a pair of distant particles which had interacted in the past, it was presumed that the measurable properties of each particle can be described independently of whatever occurs to the other particle.  However, in the 1960's, John Bell demonstrated that the underlying assumptions of EPR are incompatible: the predictions of QM cannot be reproduced by any locally causal description~\cite{bell2004}.

Bell's achievement represents an important advance, and the present work largely follows his methods.  In particular, the term ``beables'' is used for the hidden variables which are localized in space and in time and conform to either deterministic or stochastic rules, as in Bell's ``The theory of local beables''~\cite{bell1976b}.  With this term, the focus is on the rules which govern the hidden variables, the beables.  The question whether or not the beables correspond to ``something which is really there''~\cite{bell1984}, an issue which cannot be settled mathematically, is avoided at this stage--- first, one must find out which types of rules are applicable to a theory which is required to reproduce the results of QM, a reformulation~\cite{mermin1993} (for example, the Kochen--Specker theorem rules out non-contextual hidden variables \mbox{or beables).}

\begin{figure}[tb]
    \centering
    \includegraphics[scale=1.05]{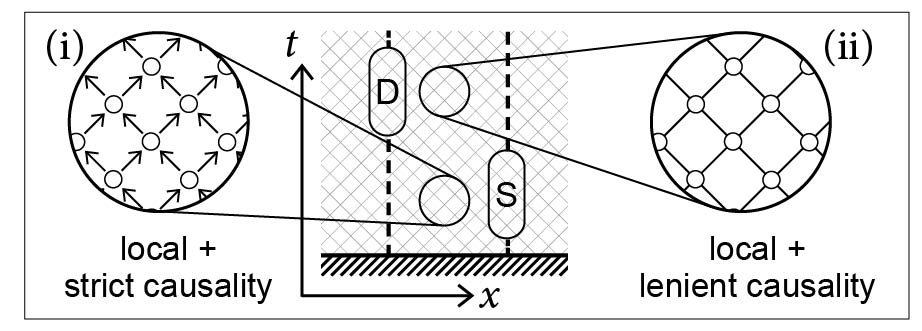}
\caption{A sketch of a spacetime region to be described by mathematical models (\textbf{{center panel}}), with marks for a source (S), a detector (D) and initial conditions (hatched).  The blowups indicate two~types of models, two possibilities for the relations between the internal variables: (\textbf{{i}}) a locally causal description, with arrows indicating probabilistic dependencies (Bayesian network); reproducing QM with such local beables is ruled out by Bell's theorem.  (\textbf{{ii}}) A time-reversal symmetric description in terms of spacetime-local beables, as per Equation~\eqref{eq:P_I(O,U)} below (Gibbs field); the symmetry is broken only by the initial conditions.  This option is not ruled out, and is the focus of the present study.  It~represents a win-win resolution of the paradox: it retains locality and avoids breaking time-reversal symmetry at every point.}
\label{fig:blowups}
\end{figure}

The present work departs from Bell's approach on an essential issue.  Even when faced with his impossibility proof, Bell still took for granted one of the assumptions: a strict form of the causal arrow of time.  Figure~\ref{fig:blowups} schematically contrasts Bell's local beables, which adhere to the strict arrow of time, with the present approach of ``spacetime-local beables,'' which are subject to local time-reversal-symmetric rules, with the symmetry broken only by imposing initial conditions, representing a ``past hypothesis.''

Such locally mediated models are not excluded from reproducing the results of QM, and over the past couple of decades concrete toy-model examples were developed to demonstrate how the correlations considered by Bell can indeed be reproduced; see~\cite{wharton2020} for a review.  In order to do so, the toy models must violate Bell's assumption of local causality, but they do not directly violate the locality condition.  Instead, they violate Bell's strict notion of causality: the hidden variables are influenced by inputs from later times.

On the other hand, as these toy models reproduce the results of QM, they must conform to signal causality.  This combination of violating strict causality while maintaining signal causality has been called lenient causality~\cite{argaman2018} (in analogy, ``lenient locality'' could be used to describe Schroedinger-picture QM: the use of a many-body wavefunction violates Bell's locality condition, but signal locality is maintained).

For reference, consider the pilot-wave theory of de Broglie and Bohm~\cite{bohm1952} (which was also important in motivating Bell~\cite{bell1982}).
In this reformulation of QM, particle positions are added to the standard description and are carried along by the Schroedinger wavefunction, resulting in a deterministic, nonlocal and strictly-forward-in-time theory which reproduces the predictions of non-relativistic QM.   The particle positions may be thought of as hidden variables representing a deeper-level description, but the wavefunction must be retained as it plays a crucial role in guiding the paths of the particles.  In other words, the particle positions are beables, but they are not local beables.

The alternative of relaxing the overly strict arrow-of-time assumption while maintaining an appropriate form of locality has not yet been developed into a reformulation of QM, but it does have a long history.  That the arrow of time is relevant is clear already in Bohr's reply to EPR, which transforms the setup into that of a delayed-choice measurement~\cite{bohr1935}.  The idea has been discussed at a conceptual or interpretational level since the 1950's, most notably by Olivier Costa de Beauregard (see, e.g.,~\cite{costa1953}), John Cramer (e.g.,~\cite{cramer1986}) and Huw Price (e.g.,~\cite{price1997}).
Note that an all-at-once point of view of microscopic physics is required here, as when considering the stationary-action reformulation of classical mechanics.

The next level of development involves mathematical toy models, limited models which reproduce the results of QM in a restricted domain.  Such toy models were only developed much more recently, at first for a single particle~\cite{schulman1997}, then for the fully entangled two-particle correlations considered by Bell ~\cite{wharton2014,wharton2020}, and only in the present decade for three-particle and $N$-particle entanglement of the so-called GHZ type~\cite{neder2024}.

The search for toy models applicable to more-complicated cases of entanglement (e.g.,~partial entanglement), {comprises a promising path toward a full locally mediated reformulation of QM}, but the present work {takes a different approach.  It sets up a general framework for such models or theories, and identifies the different types of states within them.  The framework is based on Gibbs fields, allowing for both deterministic and stochastic relations between the beables; the only additional requirement is spacetime-locality, in the sense of blowup (ii) of the figure.}

The overall state of a system is represented by the values of all the beables in an all-at-once fashion, over all times considered.  Regarding the state of the system at a specific time $t$, it is necessary to distinguish between the subset of beables pertaining to the time $t$, the ontic state (`ont' translates to the verb `to be'), and the epistemic state, the state of knowledge that the theory could allow an external observer to possess at a time $t$.  While the ontic state is stochastic, with a distribution which depends on inputs from both the past and the future, the epistemic state is a mathematical construct which necessarily depends only on the inputs and the outcomes up to the time $t$.  This state of knowledge is comparable to the quantum state.  {(It is emphasized again that although the terms ontic and epistemic are used here, the discussion is limited to the mathematical models; claiming that the ``ontic beables'' correspond to reality or that the ``epistemic state'' corresponds to knowledge would be premature.)}

Section~\ref{sec:Gibbs} introduces the mathematical modeling framework. 
Section~\ref{sec:the_state} presents the ontic and epistemic states, explaining how the quantum state $\Psi$ can be an epistemic state, the Pusey--Barrett--Rudolph theorem notwithstanding.  Section~\ref{sec4} summarizes and discusses the results, and includes a toy model example which clarifies the character of the ontic and the epistemic states.

\section{Spacetime-Local Gibbs-Field Models}
\label{sec:Gibbs}

This section presents the general framework of spacetime-local Gibbs-field (SLGF) models or theories which is to be considered here {(this framework was} briefly introduced in Ref.~\cite{neder2024}{; see also Ref.~\cite{wharton2014})}.  {It encompasses all mathematical-physics models which conform to the ``continuous action'' locality condition of Ref.~\cite{wharton2020}.  This terminology refers to Einstein's requirement of no ``spooky action at a distance,'' i.e.,~that a model which exhibits correlations between A and B includes a description of the fields (or whatever it is) in space and in time which mediate these correlations, continuously connecting between A and B.  The continuous action condition generalizes Bell's screening condition~\cite{bell2004} to situations which do not necessarily assume the strict arrow of time.  The motivation for choosing this condition, from Bell's theorem, is adequately covered in the literature, and only the points necessary for completeness are briefly repeated here.}

The SLGF framework is quite general, and includes not only the lenient-causality models and theories which could serve as locally mediated reformulations of QM, but also classical models and the locally causal models considered by Bell.
In order to lead to the ontic and epistemic states of the next section,  the presentation here is mathematical, but no attempt at formal rigor is made.  The first subsection identifies the variables involved and defines lenient causality, the second provides the expression for the probability distribution, and the third clarifies the connections to other approaches.

\subsection{The Setup}

Following Bell~\cite{bell1976b}, the discussion here relates to a specific setup consisting of several elements---sources, detectors, polarizers, beam-splitters, etc., with the location of each element fixed in space and in time as in Figure~\ref{fig:blowups}.  The input variables or settings specify further controllable details, such as the orientation of a device (e.g.,~the angle of a polarizer) or the type of measurement considered, and the set of all input variables is denoted by $I$.  The set of all outcome variables is denoted by $O$, with each {observable} outcome associated with a specific detector, clearly located in space and time.

According to the Born rule of QM, the outcomes are random variables, and the predictions for a specific problem are denoted by the (joint) probability distribution $P^{\rm QM}_I(O)$ (given the inputs, QM gives the whole distribution as an output; the outcomes are the results of the measurements{, i.e.,~the eigenvalues of the Hermitian operators, the observables corresponding to the detectors}). Applications of QM which do not fit in such a spatiotemporal setting, such as the prediction of atomic spectra {(functions of frequency which do not necessarily involve localized sources and detectors)}, are not included in the discussion here; a generally valid reformulation of QM would have to reproduce its results both in the prepare-and-measure context discussed here and in the context of characteristic frequencies which is not.

As the inputs $I$ of such a setup are experimentally controllable and the outcomes $O$ are observable, the probabilities $P_I(O)$ specify the empirically verifiable predictions of a theory.   The use of probability distributions does not restrict the type of models considered: for an outcome $A \in O$ which is completely fixed by the inputs $I$, the distribution $P_I(A)$ is \mbox{a $\delta$-function.}

One may look at a particular outcome $O_{\rm D} \in O$ of a detector D acting at time $t_{\rm D}$, and ignore the other outcomes by considering the marginal probability distribution $P_I(O_{\rm D})$.  Signal causality requires that this distribution is independent of those inputs in $I$ which pertain to times later than $t_{\rm D}$.  Formally, denoting by $I_{\leq t_{\rm D}} \subset I$ the set of inputs up to the time $t_{\rm D}$, and by $I_{>t_{\rm D}}$ the remaining inputs pertaining to later times, the requirement is that there exists a distribution $P_{I\leq t_{\rm D}}(O_{\rm D})$ which is equal to $P_I(O_{\rm D})$ in all cases, i.e.,~$O_{\rm D}$ is independent of the remaining inputs in $I_{>t_{\rm D}}$. {(A detector may act over a time interval of length} $\Delta t_{\rm D}$; this definition of signal causality uses $t_{\rm D}$ to denote the latest moment of this interval.  See below for more detailed definitions.)

Signal locality is defined similarly, replacing $I_{\leq t_{\rm D}}$ with $I_{\wedge\rm D}$, the set of all inputs within or on the past lightcone of D.  A more general statement of signal causality is that the joint distribution of all outcomes up to a time $t$ is independent of all inputs relating to \mbox{{later times,}}
\begin{equation} \label{eq:SigCaus}
    P_I(O_{\leq t})=P_{I\leq t}(O_{\leq t}) \; .
\end{equation}

While these conditions do not require any additional variables, a physical theory may also employ internal variables in its formulation, e.g.,~to describe the configuration of the system inbetween the sources and the detectors.  Denoting these variables by $U$, the full predictions of the theory are specified by the distribution $P_I(O,U)$.   The predictions $P_I(O)$ are marginals of the full distribution.  For a reformulation of
QM, these predictions would be identical to $P^{\rm QM}_I(O)$ for each particular setup, each configuration of sources, detectors and other elements in space and in time.

In a local formulation, $U$ is a set of local internal variables $u_i$, each associated with a point $x_i = ({\bf x}_i,t_i)$ of a mesh covering the relevant spacetime region. (For continuous fields one may take each $u_i$ to describe a limited region in spacetime, rather than a point, requiring agreement between neighboring $u_i$s.  As with Bell's beables, the $u_i$s need not be scalars, and may be functions or more-complicated structures.  The present discussion is not sensitive to these details.)  These variables are subject to local rules as detailed in the next subsection.  A local theory may also refer to nonlocal variables, such as the total energy integrated over space (to use one of Bell's examples), but the local rules of the complete theory suffice to determine the distribution of the local variables.  The global or nonlocal variables may be useful in formulating shortcuts for predictions in certain cases, but in a local theory they can also be omitted, as is \mbox{done here.}

The term beables refers to all the localized variables involved, $I \cup O \cup U$.  Even in a reformulation of QM with no localized hidden variables, $U \! = \! \varnothing$, the inputs and the outcomes---those which are controllable and/or observable according to QM---are beables.  They are localized in spacetime even if they follow nonlocal rules.

Bell's theorem~\cite{bell2004} concerns the locally causal models depicted in the first blowup of Figure~\ref{fig:blowups}.  For each node $i$ in such models, the probability distribution of $u_i$ is completely determined by the values of all $u_j$ in its immediate vicinity and at times slightly earlier (i.e.,~the times associated with these nodes obey $t_j < t_i$), as indicated by the arrows in the figure.  If inputs and outcomes exist at $x_i$, they too play a role.  Zooming out, Bell's strict causality requires the values of all the variables up to a time $t$, including the internal ones, to be independent of inputs at later times: 
\begin{equation} \label{eq:BstrCausality}
    P_I(O_{\leq t},U_{\leq t}) = P_{I\leq t}(O_{\leq t},U_{\leq t})
\end{equation}
(this is called ``no future-input dependence'' in Ref.~\cite{wharton2020}).  Given the initial conditions, such a model can be developed systematically forward in time, from the past to the future.

For~example, a model with just two variables, $u_1$ and $u_2$, associated with times $t_1 < t_2$, can always be described from the past to the future: first select a value of $u_1$ from the marginal distribution $P(u_1)$, and then a value of $u_2$ according to the conditional distribution $P(u_2|u_1)$.  This corresponds to a Bayesian network~\cite{koski2011bayesian}, represented graphically by a single arrow from $t_1$ to $t_2$.  In the presence of an input $I_2$ at the later time $t_2$ (a third variable), if the marginal $P_I(u_1)$ depends on $I_2$, i.e.,~when the future intervention affects the internal variables of the past, this Bayesian past-to-future recasting of the model fails.

The implication of Bell's theorem is that a locally mediated or SLGF reformulation of QM will have to violate the causality condition on the internal variables, Equation~(\ref{eq:BstrCausality}), but as QM is to be reproduced, signal causality,  Equation~(\ref{eq:SigCaus}), is necessarily obeyed.  Thus, such a model or theory must exhibit lenient causality.

\subsection{The Defining Equations}

A sufficiently general formulation of the possibilities for local models is given by the Gibbs random fields depicted in the second blowup of Figure~\ref{fig:blowups}.  The model's probability distribution is expressed as
\begin{equation} \label{eq:P_I(O,U)}
    P_I(O,U) =  \frac{1}{Z_I} \prod_l w_l(I_l,O_l,U_l) 
\end{equation}
in terms of positive local weights $w_l$.
The weights depend exclusively on local variables: each locality $l$ is associated with a short-ranged subset $X_l$ of the points $x_i$ (one may define ``short-ranged'' by limiting all the distances and time intervals between points in $X_l$ by some small constants $dx$ and $dt$).  Each of the sets $X_l$ may consist of a point $x_i$ and all its nearest neighbors, but this is not required: a set $X_l$ need not be centered on a specific $x_i$, and next-nearest neighbors may be included in $X_l$ (Figure~\ref{fig:blowups} shows an example with only nearest neighbors{; a sketch of another example appears in the next section}).

The notation $U_l$ refers to the subset of internal variables $U_l \subset U$ corresponding to the spacetime locations in $X_l$, i.e.,~$U_l = \{ u_i| x_i \in X_l \}$.  In contrast, the inputs and outcomes are here associated directly with the localities $X_l$, so that $O_l \in O$ and $U_l \in U$ are individual elements rather than subsets.  This choice allows for each input and each outcome to act either on a single point of the mesh (with $X_l = \{ x_i \}$) or over a small range, depending on the corresponding $X_l$, and will be convenient for the definitions of the next section.

In the simple cases where $X_l$ does not involve any of the sources, detectors and other elements of the setup, $I_l$ and $O_l$ are empty sets, so that $w_l(I_l,O_l,U_l) = w_l(U_l)$.
{The sets $X_l$ overlap, with the individual beables $u_i$ mediating the correlations between distant inputs and outcomes (as an example, consider the Ising model which displays long-range-ordered ferromagnetism based on nearest-neighbor interactions and weights).}

The distribution is normalized by dividing by the overall weight
\begin{equation} \label{eq:Z}
        Z_I = \sum_O \int dU \left( \prod_l w_l \right) ,
\end{equation}
where the sum and integral are over all the values of all the variables in $O$ and in $U$, with summation implied for discrete variables and integration for continuous variables.  The predictions for the outcomes are obtained by integrating out all the internal variables $U$,
\begin{equation} \label{eq:P_I(O)}
    P_I(O) = \int dU \; P_I(O,U) \, ,
\end{equation}
and are to be compared to the predictions of QM for the same setup.

While the local weights $w_l$ are taken to be positive, they may approach zero through a dependence on a global parameter.  Indeed, the Schulman toy-model, used as a primary example in Ref.~\cite{wharton2020} and in the discussion below, employs a highly peaked Lorentzian function for the $w_l\,$s, with QM reproduced only in the limit of vanishing width, $\gamma \to 0$.  The SLGF framework is thus sufficiently general to reproduce even classical dynamics, with the $w_l$ weights becoming $\delta$-functions in the appropriate limit (classical fields can be described directly by the $u_i$ beables; classical particles require suitably generalized local beables, with each $u_i$ describing the number of particles passing within $dx$ and $dt$ of $x_i$ and their properties).  {An SLGF model which conforms to Bell's local causality} can also be described as a Bayesian network~\cite{koski2011bayesian} {(see Eq.~(\ref{eq:BstrCausality}) and Figure~\ref{fig:blowups} above)}.

Several additional remarks are in order.  These models are quite analogous to those of statistical physics, though they range over spacetime rather than just over space.
In fact, the terminology ``Gibbs field,'' which refers to any positive probability distribution of any number of stochastic variables, is borrowed from statistical physics (such distributions can always be written in terms of positive weights as above).
In that context, the positive local weights $w_l$ are expressed as exponentials, $w_l = \exp(-\beta \phi_l)$, of local potentials $\phi_l$ multiplied by an inverse temperature $\beta$.  The ``partition sum'' $Z_I$ is global, even though the $w_l \,$s are local; this is dictated by the mathematical rules of probability theory.

The spacetime-locality of the $w_l$ weights is mirrored in the dependencies of the model variables $u_i$.  Specifically, the conditional probability of $u_i$, given the values of all the neighboring $u_j$s and all the corresponding $O_l\,$s, is independent of any of the non-neighboring variables~\cite{pearl1988}.  Neighboring here refers to situations with $x_i$ and $x_j$ appearing in the same $X_l$; for the outcomes $O_l$, it simply refers to localities which include $x_i$.  Independence refers to statistical independence for the non-neighboring variables in $U$ and in $O$, and to functional independence for the distant inputs $I_{l'}$ for which $x_i \notin X_{l'}$.

The discretized mesh over spacetime is introduced here merely for convenience; the requirement that the $w_l$ weights be short ranged corresponds directly to the continuous action condition, which is formulated for a continuum ~\cite{wharton2020} (due to the discrete mesh here, the thickness of the screening regions must be taken larger than the range of the \mbox{$w_l\,$ weights).}

Quantum field theory employs an expression similar to Equation~(\ref{eq:P_I(O,U)}) for the probability amplitude of the set of outcomes $O$, given the inputs $I$.  This type of theory is spacetime-local in the sense that it can be expressed using local weights, but it is not a Gibbs-field theory as the $w_l$ weights are complex phase factors rather than being positive, and the probability amplitude is squared to give the probability.  The configurations corresponding to different sets of values of the $u_i$ variables are all superposed, and it is not possible to identify probabilities for each configuration.

It is natural to require the weights $w_l(U_l)$ for the bulk of spacetime to be time-reversal symmetric, so that each of the variables $u_i$ is influenced by other variables associated with both earlier and later times.  The time-reversal symmetry may then be broken by the imposition of initial conditions and/or by special rules for the weights $w_l(I_l,O_l,U_l)$ applicable to detectors, where dissipative effects are expected.  Indeed, only SLGF models for which time-reversal symmetry is broken and there is no retro-signaling can reproduce the results of QM, as that requires signal causality, i.e.,~an asymmetric causal arrow of time.

The present work does not specify the $w_l$ weights; instead, it reports a {step on the route toward} reproducing QM in terms of such SLGF models, the identification of ontic and epistemic states at time $t$  (the topic of Section~\ref{sec:the_state} below).  The discussion is qualitative in this sense, and the results are broad conclusions rather than narrow examples.

\subsection{Other Stochastic Approaches}

This subsection clarifies the relation between the present work and a number of existing approaches.  It is not required for the developments below, and can be skipped on a first reading.

\textbf{{Markov fields:}}
Markov random fields are stochastic models slightly more general than Gibbs fields in that they allow for vanishing weights, which can lead to complications~\cite{wiki2025markov_field} {(for an example, see Section~\ref{sec4dot2} below)}. 
A separate issue is that, for physicists, they might bring to mind Markov processes which evolve in an orderly manner from the past to the future, in stark contrast to the lenient causality models considered here.  For these reasons, the terminology of Markov random fields is avoided here (vanishing weights are included as a limiting case of small positive weights).

\textbf{{Markov blankets:}} The term minimal Markov blanket~\cite{pearl1988} refers to the aforementioned set of neighbors of a variable $u_i$, the union of $\{ u_j | \, j \neq i \;  \land  \; \exists_l \, x_i,x_j \in X_l \}$ with $\{ O_l | \, x_i \in X_l \}$ and $\{ I_l | \, x_i \in X_l \}$ (for an outcome $O_l$, the minimal Markov blanket is simply $\{ u_i | \, x_i \in X_l \}$).  The idea is that the conditional probability distribution for $u_i$ is given directly by the $w_l$ weights in which $u_i$ participates and the values of the variables participating in them, irrespective of the values of the other variables in the model which are located further away.  The local or short-ranged nature of the weights in a{n SLGF} model may thus be characterized by noting that all the variables in the minimal Markov blanket of $u_i$ are within $dx$ and $dt$ of $x_i$.

Over the decades, there have been several attempts to reproduce the results of QM with stochastic models.  Some of these have achieved remarkable successes, but they do not result in a{n SLGF} theory or reformulation.

\textbf{{Stochastic mechanics:}} This approach~\cite{nelson1966,nelson2012} begins with the assumption that a particle of mass $m$ follows a time-reversal-symmetric diffusive path with a diffusion coefficient of $\hbar/m$, subject to a suitably defined acceleration $a = F/m$.  The analysis deduces that a pair of velocity fields (material velocity and osmotic velocity) guide the particles, similar to the pilot wave theory, and yields two coupled differential equations for them.  These equations can be combined into a single complex equation, reproducing the Schroedinger equation.

However, the $F = ma$ rule is applied to the guiding fields, i.e.,~to the ensemble of paths, rather than to an individual path which is a realization of a specific instance within such a model.  It thus violates the continuous-action locality condition~\cite{wharton2020}, i.e.,~the model is not a{n SLGF} model.  Indeed, when two particles are considered, the resulting dynamics is manifestly nonlocal: the Schroedinger equation describes evolution in the two-particle configuration space{, rather than in space and time}.

\textbf{{Stochastic quantization:}}
Uses a Gibbs field to reproduce quantum field theory in Euclidean space, but when Wick rotation is invoked in order to apply the approach to Minkowski space, the resulting weights are complex, and the result is not a{n SLGF} model~\cite{damgaard1987}.

\textbf{{Pilot wave theory:}}
Although it is customary to think of the approach of de Broglie and Bohm~\cite{bohm1952} as a deterministic theory, it can also be viewed as a stochastic model---the predictions of QM are only reproduced if the initial particle positions are distributed according to the Born rule.  The guiding equations are nonlocal.
 
\textbf{{Stochastic electrodynamics:}}
A number of quantum phenomena can be reproduced in an entirely classical theory by considering a classical particle in a fluctuating electromagnetic field, with initial conditions of energy $\frac{1}{2}\hbar\omega$ per mode~\cite{delapena1996,delapena2014,boyer2019}.  However, as local causality is retained, Bell-inequality-violating correlations cannot be reproduced~\cite{marshall1989}.

\textbf{{Indivisible stochastic processes:}}
A recent development which casts quantum processes as stochastic processes, Ref.~\cite{barandes2025}, does so by introducing nonlocality in time (in addition to spatial nonlocality), while keeping the strict causal arrow of time.

\textbf{{Forward-backward simulations:}}
The $Q$-based approach developed by Drummond and Reid, see Ref.~\cite{reid2026} and references therein, is a stochastic simulation process for quantum phenomena {which displays lenient causality but does not restore locality in the context of Bell correlations}.

\textbf{{Weak values:}}
Sutherland has suggested that so-called weak values may serve as spacetime-local beables, {in particular for quantum circuits}, see~\cite{wharton2024} and references therein.  The weak values behave locally in the sense that the value for a certain qubit is undisturbed by a gate acting on other qubits, but it has not been shown that they conform to the continuous action condition for a general circuit, i.e.,~that they result in an SLGF model.

\section{The State of the System at Time \boldmath{$t$}}
\label{sec:the_state}

When considering a{n SLGF} model and focusing on a time $t$, it is natural to take the values of the variables $u_i$ for which $t_i \simeq t$ as representing the internal state of the system at that time.  This set of variables, denoted $U_t$, is the ontic state of the model at $t$.  The~definition and use of $U_t$ is the topic of the first subsection below.

What knowledge a user of such a model or theory can have at time $t$ is a different question. This epistemic state is the topic of the second subsection, though it appears already in the equations of the first.  It shares many features with the quantum state $\Psi$ \mbox{(or $\hat\rho$),} and in particular, the epistemic state changes abruptly between times right before and right after a measurement, in close analogy to ``collapse'' of the wavefunction.

\subsection{The Ontic State}

The full predictions of any SLGF model are given by the full distribution $P_I(O,U)$, obtained from its local weights $w_l$ according to Equation~(\ref{eq:P_I(O,U)}). In addition to looking at the empirical predictions, $P_I(O)$, i.e.,~the marginal distribution obtained by integrating out (or~summing over) all of the internal variables in $U$, one may retain a subset of the internal variables, and in particular, the subset $U_t$ associated with a time $t$.

\begin{figure}[b]
    \centering
    \adjustbox{trim={.05\width} {.12\height} {0.1\width} {.1\height},clip}%
    {\includegraphics[scale=0.4]{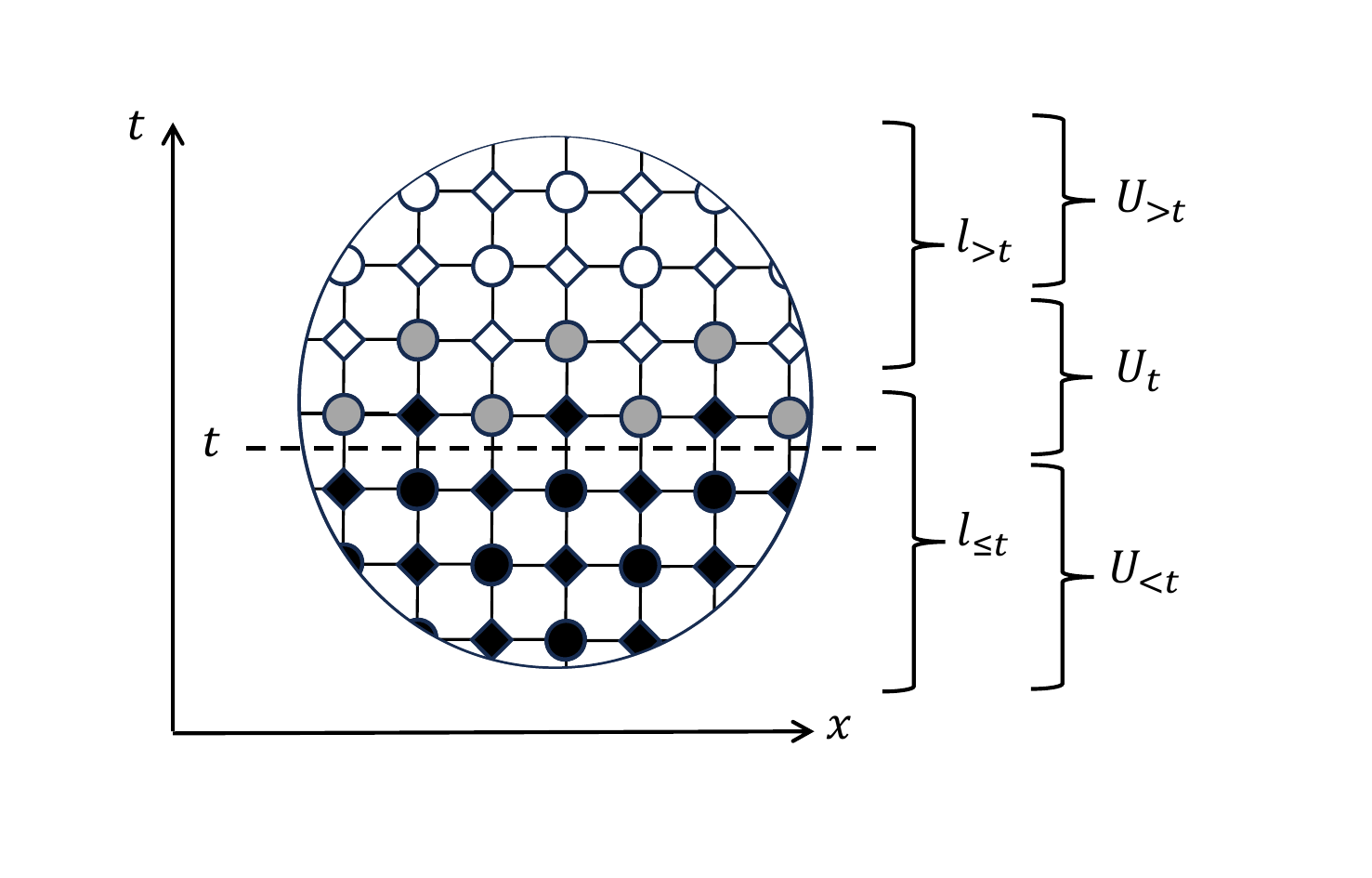}}
\caption{A spacetime lattice of sites $i$, depicted by circles, with localities $l$ depicted by diamonds.  A time $t$ is indicated by the dashed line.  The localities are divided by the line into two groups, $l_{\leq t}$ (black) and $l_{>t}$ (white), while the sites are divided into three, $U_{<t}$ (black), $U_t$ (gray) and $U_{>t}$ (white).}
\label{fig:ontic}
\end{figure}

In order to define $U_t$, the set of all local weights $w_l$ is partitioned into two disjoint sets pertaining to times up to $t$ and to times after $t$, see Figure~\ref{fig:ontic}, where the localities $X_l$ are displayed explicitly (each site $i$ has 4 neighboring localities $l$ and 8 neighboring sites~$j$).  Referring to the indices $l$ of the weights $w_l$, the future set, $l_{>t}$, contains all $l$s for which $\forall_{x_i \in X_l} \; t_i > t$, i.e.,~$w_l$ acts only at times after $t$.  The set, $l_{\leq t}$, relating to the past and the present of $t$, contains all the remaining indices $l$ (it refers to both weights $w_l$ which are wholly to the past of $t$,  and to weights which straddle $t$).

The set $U_t$ contains all internal variables $u_i$ which participate in weights of both types,~i.e.,
\begin{equation}
    U_t = \{ u_i | (\exists_{l \in l_{\leq t}}\; x_i \in X_l) \land  (\exists_{l \in l_{>t}}\; x_i \in X_l) \}  \, .
\end{equation}
The remaining internal variables can then be partitioned into earlier and later times,  $U_{<t}, U_{>t} \subset U$, with $U_{<t} = \{ u_i | \, x_i \not\in \cup_{l \in l_{>t}} X_l \}$, etc.  As the inputs and outcomes have been taken to act directly on the weights $w_l$, the sets $I$ and $O$ are partitioned into two disjoint sets each, $I_{\leq t}, I_{>t}$ and $O_{\leq t}, O_{>t}\,$, in direct correspondence to $l_{\leq t}, l_{>t}\,$.
 
With this notation, the marginal distribution for $U_t$, the system's internal variables at time $t$, in addition to the outcomes $O$, is
\begin{equation} \label{eq:P_I(O,U_t)}
P_I(O,U_t) = \frac{1}{Z_I} w_{\leq t}(I_{\leq t},O_{\leq t},U_t) \, w_{>t}(I_{>t},O_{>t},U_t)  \, ,
\end{equation}
where $w_{\leq t} = \int dU_{< t} \;\prod_{l \in l_{\leq t}} w_l$ contains the product of the weights up to the present $t$, and $w_{>t} = \int dU_{>t} \;\prod_{l \in l_{>t}} w_l$ represents the future.  While the complexity of the ontic state grows linearly with the number of elements $u_i$ in $U_t$, the complexity of the probability distribution $P_I(O,U_t)$ grows exponentially in cases with nontrivial correlations, as it specifies the odds for each combination of values of the individual $u_i$s.

The normalization factor, Equation~(\ref{eq:Z}), can also be written as
\begin{equation} \label{eq:Z_I(O,U_t)}
Z_I = \sum_O \int dU_t \, w_{\leq t}(I_{\leq t},O_{\leq t},U_t) \, w_{>t}(I_{>t},O_{>t},U_t)  \, ,
\end{equation}
where again $\int dU_t$ implies integration over all the variables in $U_t$.  The predictions of the model are reproduced from this marginal by direct integration,
\begin{equation} \label{eq:P_I(O)t}
    P_I(O) = \int dU_t \; P_I(O,U_t) \, .
\end{equation}

{If $O_{\leq t}$, the outcomes up to the time $t$ are considered known, the conditional probability is}
\begin{equation} \label{eq:P_I(O>,U_t|O<)}
    P_I(O_{>t},U_t | O_{\leq t}) =  \frac{P_I(O,U_t)}{P_I(O_{\leq t})} \, .
\end{equation}
{One} may also write this conditional probability just as in Equation~(\ref{eq:P_I(O,U_t)}), replacing $Z_I$ in the denominator by 
\begin{equation} \label{eq:Z_I|Oleq}
Z_{I|O_{\leq t}} = \sum_{O_{>t}} \int dU_t \, w_{\leq t}(I_{\leq t},O_{\leq t},U_t) \, w_{>t}(I_{>t},O_{>t},U_t)  \, ,
\end{equation}
as the marginal probability $P_I(O_{\leq t})$ is just the ratio of the two normalization factors,
\begin{equation}
    P_I(O_{\leq t}) = \frac{ Z_{I|O_{\leq t}} } {Z_I} \, .
\end{equation}
{Note} that signal causality, Equation~(\ref{eq:SigCaus}), requires this ratio to be independent of $I_{>t} \subset I$, a remarkable requirement, but the expressions here do not assume this.

Clearly, one may consider other types of marginal and conditional probabilities, e.g.,~the marginal distribution of $U_t$ given the outcomes up to $t$, i.e.,~$P_I(U_t|O_{\leq t})$, and disregarding the future outcomes $O_{>t}$.  It is obtained from Equation~(\ref{eq:P_I(O>,U_t|O<)}) by summing over $O_{>t}$.  Similarly, predictions for the future outcomes $O_{>t}$ given the past and present outcomes $O_{\leq t}$ and disregarding the state at $t$ are obtained by integrating over $U_t$,
\begin{equation} \label{eq:P_I(O>|O<)}
     P_I(O_{>t} | O_{\leq t}) =  \int dU_t \; P_I(O_{>t},U_t | O_{\leq t})  \, ,
\end{equation}
as in Equation~(\ref{eq:P_I(O)t}) above.

\subsection{The Epistemic State}
\label{sec:epistemic}

When considering Equation~(\ref{eq:P_I(O,U_t)}) and its consequences, it is natural to identify the integrated weight $ w_{\leq t}(I_{\leq t},O_{\leq t},U_t)$ as an epistemic state, representing the information about $U_t$ which may be available to a user of the theory up to the time $t$.  It obviously represents partial knowledge regarding $U_t$; the actual value of $U_t$ fluctuates and may not be accessible, and in order to merely obtain its distribution, one also requires the integrated future weight, $ w_{>t}(I_{>t},O_{>t},U_t)$.

The correspondence between these integrated $w$ functions and those of QM is quite remarkable.  The quantum state of a system at time $t$, the wavefunction $\Psi_t$ or the probability-density operator $\hat \rho_t$, depends on the types of measurements performed up to $t$ and on their outcomes.  It also depends on internal variables, in the sense that the values of the quantum state depend on the Hilbert-space basis used for its representation.  It is thus analogous to the epistemic state $ w_{\leq t}(I_{\leq t},O_{\leq t},U_t)$.

In order to obtain predictions for measurements, Born's rule multiplies this state by the projection operator for a specific outcome or set of outcomes, $\hat\Pi_{O_{>t}}$, and takes the trace:
\begin{equation} \label{eq:BornRule}
    P^{\rm QM}_I(O_{>t}|O_{\leq t}) = \Tr{\hat \rho_t \hat \Pi_{O_{>t}}}  \; .
\end{equation}
{The} projection operator $\hat\Pi_{O_{>t}}$ depends on the types of measurements performed at or after $t$, and on the outcomes and the basis, similar to the future integrated weight $ w_{>t}(I_{>t},O_{>t},U_t)$.  For measurements made after $t$, the projection operator $\Pi_{O_{>t}}$ is the operator corresponding to the measured observable, unitarily evolved back in time to $t$.  The trace corresponds to the integration in Equation~(\ref{eq:P_I(O>|O<)}).

This correspondence leads to striking similarities between the quantum formalism and the present SLGF models, but there are also differences.  One difference is the appearance of the normalization factor in Equation~(\ref{eq:P_I(O,U_t)}), whereas in QM the density matrix $\hat \rho$ and the projection operators $\hat \Pi$ are separately normalized ($\Tr{\hat\rho} = 1$ and $\sum_O \hat\Pi_O = \hat I$).  We will return to this point in the discussion below.

The similarities are:
\begin{itemize}
    \item If no inputs and outcomes are involved between times $t_1$ and $t_2$, the ``evolution'' of the past (and future) {integrated} weights, i.e.,~the change in $ w_{\leq t}$ (as well as $w_{>t}$) over this time interval, conserves the relevant information in the following sense: Equation~(\ref{eq:P_I(O>|O<)}) holds with the same partition between conditioning and conditioned variables, regardless of whether the integration is over $U_{t_1}$ or $U_{t_2}$.  This conservation of information parallels unitary evolution in QM.
    \item If a measurement occurs between $t_1$ and $t_2$, there is an abrupt change in the epistemic state $w_{\leq t}$.  Its value after the measurement depends on what measurement was performed and what outcome was obtained, in the same manner as for collapse of the wavefunction in QM.
    \item The epistemic state $ w_{\leq t}$ relates to times up to $t$, and thus bisects spacetime into past and future on a spacelike hypersurface, just like the quantum state.  This coordinate dependence, together with the update of the epistemic state over all space due to a measurement in a particular region within space, might be interpreted as generating a tension with special relativity.  In contrast, the underlying {SLGF} model does not require a hypersurface of simultaneity.
    \item When many degrees of freedom are involved and the number of components in the ontic state $U_t$ is large, the complexity of the epistemic state generically grows exponentially, for both $w_{\leq t}$ and for $\Psi$.  {(The argument here concerns the exponential number of combinations of  all the} $u_i$ involved, $\sim \exp(|U_t|)$, as mentioned above; the complexity of many-particle epistemic states can also be argued based on the combinations of possible measurements the particles may eventually undergo, \mbox{$\sim\exp(|I|)$,} each specified by its measurement setting in $I$~\cite{wharton2014}.)
    \item Whereas the {integrated} weights $w_{\leq t}$ and $ w_{>t}$ give direct information on the {ontic} state $U_t$ of the underlying {model (in contrast to the abstract nature of the quantum state)} the information regarding the empirical predictions is obtained by integrating over $U_t$, cf.\ Equation~(\ref{eq:P_I(O)t}).  One may thus transform from $U_t$ to alternative integration variables, while keeping the predictions unchanged.  This is closely analogous to transforming $\Psi$ from one basis of Hilbert space to another.
\end{itemize}

{The conclusion} is that the {SLGF} framework, by its very structure, naturally provides an explanation of all these features of the quantum formalism. This is seen here to hold regardless of whether the predictions of QM are reproduced, and even regardless of \mbox{signal causality.}

The implication that the quantum state is an epistemic state ostensibly contradicts the Pusey--Barrett--Rudolph theorem~\cite{pusey2012}, but that work tacitly assumes the strict arrow of time.  In particular, the assumption there was that ``systems that are prepared independently have independent physical states.''  Although this does not explicitly mention causality or the arrow of time, it is clear that it will not necessarily hold in a lenient-causality {SLGF} model. In particular, systems $a$ and $b$ which are prepared independently before time~$t$, will have independent epistemic states $ w_{\leq t}$ at time $t$, but if they are later measured jointly, the relevant future weights $ w_{> t}$ may well result in correlations between their ontic states, $U_t^\text{(a)}$ and $U_t^\text{(b)}$. It is thus no surprise that the conclusions of Ref.~\cite{pusey2012} do not hold (see,~e.g.,~Ref.~\cite{leifer2014}).

\section{Discussion}
\label{sec4}

A century ago, the addition of Born's rule to the quantum formalism cemented the departure of fundamental physics from ``complete causality,'' i.e.,~determinism: microscopic physics involves randomness and requires a stochastic approach.  ``Causality'' in the sense of the strict causal arrow of time, that future interventions cannot affect what ``really happens'' in the past, was ostensibly left intact.  The lenient-causality approach denies both these types of causality from the outset, keeping only signal causality, the arrow of time associated with macroscopically available information.

This is just the type of innovation one could expect following the early-20th-century revolutionary concepts of the relativity of simultaneity and of the curvature of spacetime: the influences of an external input or intervention at $t$ may have microscopic effects on adjacent times $t\!\pm\!dt$ symmetrically, both after and {\textit{before}} $t$.  In such a theory, breaking the symmetry by imposing initial conditions would be responsible for the fact that information-carrying signals can only be received after they have been generated (cf.\ Figure~\ref{fig:blowups}).

While these ideas coincide with those of the transactional interpretation~\cite{cramer1980,cramer1986,cramer2016}, the goal pursued here is {a step towards} developing a reformulation of QM, with different mathematical expressions but coinciding predictions~\cite{feynman1964}.  The requirement that a complete description must be local is adopted~\cite{EPR1935,bell2004}, with the modification of allowing for \mbox{lenient causality.} 

The following subsections summarize the results above, apply the general expressions to one of the available toy models, and present conclusions.

\subsection{Summary}

Using standard probability theory, the generalized locality requirement leads to adopting SLGF models as a framework, i.e.,~conducting a study of stochastic models which are spacetime-local, see Figure~\ref{fig:blowups}.  Indeed, all existing locally mediated toy models of quantum phenomena can be cast in these terms.  The model or theory is defined in terms of the local weights $w_l$ appearing in Equation~(\ref{eq:P_I(O,U)}), which resembles expressions of classical statistical physics (each weight factor can be thought of as the exponent of a local potential, $w_l \propto \exp(-\beta \phi_l)$, but acts over space and time rather than just over space).

The lenient-causality approach requires a block-universe stance: the system is described by all the input variables $I$, the internal variables $U$ and the outcome variables $O$ at once, regardless of which variables pertain to earlier times and which to later times (a Lagrangian rather than a Newtonian approach).

One may still focus on a time $t$ intermediate between preparations and measurements (so that no inputs and outcomes refer to $t$), and take $U_t$, the subset of internal variables pertaining to the time $t$, to represent the ontic state at $t$.  The SLGF formulation immediately gives the joint probability distribution for this state and for the outcomes, Equation~(\ref{eq:P_I(O,U_t)}), in terms of the integrated weights, $w_{\leq t}$ and $w_{>t}$, representing times up to the present $t$ and times of its future, respectively.

The past {integrated} weight $w_{\leq t}$ depends on the inputs and the outcomes up to $t$, i.e.,~on $I_{\leq t}$, $O_{\leq t}$, and the future weight $w_{>t}$ depends on those of the future, $I_{>t}$, $O_{>t}$.  The predictions $P_I(O)$ are obtained from the product of the two integrated weights as a marginal distribution, by integrating over $U_t$ and normalizing, Equations~(\ref{eq:P_I(O,U_t)})--(\ref{eq:P_I(O)t}).

The past integrated weight $w_{\leq t}$ contains all the information regarding $U_t$ which exists in the model up to the time $t$, and therefore it is natural to refer to it as the epistemic state of the description.  Its correspondence to the state $\Psi$ or $\hat \rho$ in QM implies that the quantum state too is an epistemic state.  The conclusion of the Pusey--Barrett--Rudolf theorem \cite{pusey2012} that the quantum state is not an epistemic state is not applicable, as that theorem, like Bell's, assumes a strict arrow of time.

{The predictions for future measurements are expressed in terms of the integrated weights $w_{\leq t}$ and $w_{>t}$ in an expression, Equation~(\ref{eq:P_I(O>|O<)}), which is quite analogous to the Born rule of QM, Equation~(\ref{eq:BornRule}).  The latter too expresses the predictions by combining the quantum state (analogous to $w_{\leq t}$), representing the inputs of the past, with projection operators (analogous to $w_{>t}$ ) representing measurements at either the time $t$ or its future.}

Just like unitary evolution in QM, the integrated weights $w_{\leq t}$ and $w_{>t}$ depend on the time $t$ in a manner which preserves the information regarding the predictions $P_I(O)$ over time intervals when no measurements are made.  Comparing times just before and just after a measurement, they change abruptly.  In fact, the epistemic state just after a measurement, $w_{\leq t}$, depends on the result of that measurement, just like collapse in QM.

The expressions involving $w_{\leq t}$ and $w_{>t}$ apply to any  SLGF model, whether or not the model agrees with QM (and even if it displays backward-in time signaling, which occurs in models with restrictive final conditions, e.g.,~with post-selection~\cite{aharonov2008}).

In fact, this characteristic time dependence, with abrupt changes at measurements punctuating information-preserving time dependence, is a generic feature of epistemic states, as has been repeatedly discussed in the past.  It is particularly interesting to compare the present approach with Spekkens's toy theory~\cite{spekkens2007}.  That work shows how many of the striking features of QM can be reproduced in a model in which an ``epistemic restriction'' principle is introduced, requiring the epistemic state to contain only half of the information required to identify the ontic state, even when knowledge is maximal.  This parallels Equation~(\ref{eq:P_I(O,U_t)}) above, where the information from the epistemic state, the past weights $w_{\leq t}$, must be combined with the future weights $w_{>t}$ in order to give the distribution of the ontic state.  An advantage of the present approach is that no ad hoc principles need to be introduced---the results follow directly from the structure of spacetime, cf.\ Figure~\ref{fig:blowups}.

In a broader context, many works take the strict arrow of time for granted, as though it follows from the signaling arrow of time.  In retrospect, it is clear that this is a separate assumption, and as such requires justification.  For example, the notion of a free-variable input is often taken to imply freedom from the past~\cite{bell1977}.  The very notion of something ``existing'' or ``being'' at a certain time is taken to imply independence of future inputs.  While this is clearly in accordance with normal intuition, it violates time-reversal symmetry, and should thus be treated with suspicion. The lenient-causality approach makes less assumptions than the standard approach, not more.

\subsection{Toy-Model Example\label{sec4dot2}}

As a simple concrete example, consider the single spin of the Schulman model \mbox{({Section~9.1}} of Ref.~\cite{schulman1997}).  It is described by a time-dependent angle $\theta(t)$ in a fixed plane, say, the $y,z$ plane, subject to an initial condition $\theta(0)=\theta_i$ at $t=0$, and a measurement at an angle $\theta_{\rm D}$ at time $t=T$.  The final angle $\theta(T) = \theta_{\rm f}$ is equal to either $\theta_{\rm f,1} = \theta_{\rm D}$ or $\theta_{\rm f,0} = \theta_{D}+\pi$ (mod~$2\pi$) according to whether the outcome is $O_{\rm D}=1$ or $O_{\rm D}=0$, respectively, i.e.,~whether the result of the measurement is up or down relative to the angle $\theta_{\rm D}$ (there are two inputs, $I=\{\theta_i,\theta_{\rm D}\}$, and a single outcome, $O = \{ O_{\rm D} \}$).

At intermediate times the angle $\theta(t)$ is subject to stochastic kicks, with a Cauchy (Lorentzian) distribution which is folded to obtain the $2\pi$-periodicity of the angle: the weights $w_l$ of Equation~(\ref{eq:P_I(O,U)}) are
\begin{equation}
    w_{dt}(d\theta) = \frac{1}{2\pi} \frac{\tanh(d\gamma/2)}{\sin^2(d\theta/2)+\cos^2(d\theta/2)\tanh^2(d\gamma/2)}  \, ,
\end{equation}
where $d\theta$ is the change in $\theta(t)$ over the time-step $dt$, and $d\gamma > 0$ is the width parameter for that time step, which is taken to be very small.

Convolutions of the Cauchy distribution are Cauchy distributions, with the widths adding linearly---the distribution is ``stable'' in this sense~\cite{schulman1997}.  This holds as well for the above $2\pi$-periodic Cauchy distribution.  The integrated weight for the entire time interval $(0,T)$ is thus a periodic Cauchy distribution of overall width $\gamma = \int_0^T d\gamma$.

The predictions of QM obtain in the limit of small overall width parameter, $\gamma \to 0$, for which one may use
\begin{equation} \label{eq:w_Sch}
    w(\Delta\theta) \simeq \frac{1}{\pi} \frac{\gamma}{4\sin^2(\Delta\theta/2)+\gamma^2}
\end{equation}
for the weights integrated over all $U$, with $\Delta\theta = \theta_{\rm f}-\theta_{\rm i}$.  The normalization factor of Equation~(\ref{eq:Z}) is
\begin{equation}
    Z_I \simeq \frac{1}{\pi} \frac{\gamma}{\sin^2(\Delta\theta)+\gamma^2} \, ;
\end{equation}
it is periodic with period $\pi$, so either value of $\theta_{\rm f}$ can be used.  The outcome probabilities, Equation~(\ref{eq:P_I(O)}), thus reproduce the standard result {of QM} for spin projections,
\begin{equation} \label{eq:P_Sch}
    P_I(1) = 1-P_I(0) = \cos^2(\Delta\theta_1/2)  \, ,
\end{equation}
in the $\gamma \to 0$ limit~\cite{schulman1997}, where $\Delta\theta_1$ refers to $\theta_{\rm f,1}-\theta_{\rm i}$.

The overall path of the spin, $\theta(t')$ for $t' \in [0,T]$, is a Lévy flight characterized by many small kicks of order $d\gamma$ and a single significant kick of magnitude $\simeq \Delta\theta$, which may be described as a kink in the path (if $\Delta\theta = 0$ is possible, no kink occurs).  Due to the imposition of both initial and final conditions, one can no longer state that the individual kicks are Cauchy distributed: for generic inputs a kink will necessarily occur {(the kink may be described as a collapse-like event of the ontic state}~\cite{wharton2020}, but this should not be confused with collapse of the epistemic state).
{The quantitative description of the kink in this toy model demonstrates the need to take the limit $\gamma \to 0$ carefully and avoid vanishing weights, as a naive replacement of the narrow-width Cauchy distribution by a $\delta$-function would not lead to meaningful results.}

The epistemic state for this model, $w_{\leq t}$ of Equation~(\ref{eq:P_I(O,U_t)}), is also a periodic Cauchy distribution, Equation~(\ref{eq:w_Sch}); in this case the width $\gamma$ is $\gamma_{\leq t} = \int_0^t d\gamma$, the sum of the strengths of the kicks accumulated up to $t$, and $\Delta \theta = \theta(t) - \theta_i$.  The future integrated weights $w_{>t}$ are given by the same expression, with width $\gamma_{>t} = \int_t^T d\gamma$ and $\Delta\theta = \theta_{\rm f} -\theta(t)$; the two possible values of $\theta_{\rm f}$ are as above, with the specific value selected according to the probability \mbox{of Equation~(\ref{eq:P_Sch}).}

The ontic state, $U_t = \theta(t)$ at an intermediate time $t$, is thus distributed per Equation~(\ref{eq:P_I(O,U_t)}) very near $\theta_{\rm i}$ and $\theta_{\rm f}$, in the following sense: for any small accuracy parameter $\delta > 0$, one can guarantee that with probability at least $1-\delta$ the value of $\theta(t)$ falls within $\delta$ of either $\theta_{\rm i}$ or $\theta_{\rm f}$, by choosing $\gamma$ of order $\delta^2$ or smaller (the probability of obtaining a path $\theta(t')$ with two or more kinks, i.e.,~at least two distinct non-overlapping time intervals with $|\int d\theta| > \delta$, is then smaller than $\delta$).

The probability for $\theta(t)$ falling near $\theta_{\rm i}$ (or near $\theta_{\rm f}$) depends on the ratio of the widths $\gamma_{\leq t}/\gamma_{>t}$ in a continuous manner.  The ontic state $U_t$ thus depends symmetrically on both its past and its future.

Note that although the epistemic state $w_{\leq t}$ is ostensibly a quite general (positive, integrable) function of $\theta(t)$, it is ``rigid'' in the sense that it is completely parameterized by two real parameters, $\theta_i$ and $\gamma_{\leq t}$.  It is therefore reasonable for it to correspond to the quantum state $\Psi$, which for a single spin is characterized by a single complex number (it is normalized and an overall phase can be excluded).  The same applies to the future integrated weight $w_{>t}$ and its correspondence to the projection operators $\hat\Pi_{O_{\rm D}}$ of Equation~(\ref{eq:BornRule}).  

The Schulman model has been extended to describe two entangled spins prepared in the singlet state~\cite{wharton2014,wharton2020}.  In this case the internal variables are the two paths $\theta_a(t)$ and $\theta_b(t)$, and the initial condition involves a $\delta$-function weight requiring $\theta_{a,\rm i}$ and $\theta_{b,\rm i}$ to be in opposite directions.  Measurements are applied separately to the spins $a$ and $b$, and the predictions reproduce the Bell correlations of QM in full.  A single kink occurs along the paths of both spins, allowing the model to reproduce the correlations predicted by QM (this applies also to the extension to $N$ spins in the GHZ state~\cite{neder2024}, a single kink occurs for all $N$ spins).

Note that the structure of the two expressions for the probabilities, Equation~(\ref{eq:P_I(O>|O<)}) of {an SLGF} model and Equation~(\ref{eq:BornRule}) of QM, is quite different.  The latter, the Born rule, is a trace of a product of two factors, the density matrix $\hat \rho_t$ depending only on the past inputs and outcomes, and the projection operator $\hat \Pi_{O_{>t}}$ depending only on those of the future.  In contrast, Equation~(\ref{eq:P_I(O>|O<)}) is based on Equation~(\ref{eq:P_I(O,U_t)}), and gives this probability as a ratio of two such quantities.

The structure of the Born rule has important consequences---it can be used to prove signal causality from the fact that the projection operators of any measurement add up to unity, and to prove the Tsirelson bound~\cite{cirelson1980}.  Finding the general conditions under which the expression of Equation~(\ref{eq:P_I(O>|O<)}) can be transformed into the form of the Born-rule expression, Equation~(\ref{eq:BornRule}), is thus identified as an important direction for future study.

\subsection{Conclusions}

If a locally mediated reformulation of QM, {a description of quantum phenomena within the SLGF framework}, can indeed be found, much would follow.  For example, it would have immediate consequences for recent discussions of quantum gravity, such as Refs.~\cite{oppenheim2023,bose2017}, which deal with the question of whether or not the geometry of spacetime itself can be in a superposition (with the standard strict assumptions regarding the arrow of time taken for granted).  The answer would be that all superpositions only represent a particular feature of the epistemic state of a system, not its ontic state.  Additionally, such a reformulation would be expected to provide a new starting point from which to begin developing a novel candidate for a theory of quantum gravity.

Of course, that is a big if.  As with any well-posed unsolved mathematical problem, a negative answer is also possible.  A proof that such a reformulation cannot be found, for all locally mediated models or for a subclass, would also constitute a dramatic advance in our understanding of QM: such a no-go result would represent a significant strengthening of Bell's theorem.

Returning to the present work, the identification of distinct ontic and epistemic states resolves the measurement problem (see, e.g.,~Ref.~\cite{leggett2005}) in the sense that there is no need to explain why and how measurements have a single outcome: the model is stochastic to begin with.  The Schroedinger equation is more closely analogous to the Liouville equation, providing information on where the system may go in phase space, than to the Newton equation which specifies the evolution of the ontic state of the system.

However, the inclusion of measurements in the SLGF description is distinctly artificial.   Given that measurement devices are also physical systems, how can the outcomes $O$, which presumably merely reflect possible events involving the internal variables $U$ of the measurement device, behave so differently?  In particular, how can lenient causality be brought about in a detailed model of a detector, with signal causality for the outcome coexisting with future-input dependence for the internal variables?
This problem will be tackled in a sequel to the present work.

\smallskip
\paragraph*{Acknowledgments:}
Valuable discussions with Ofek Bengyat, Eyal Buks, Marios Christodoulou, David Kutasov, Tim Maudlin, Markus Müller, Izhar Neder, Ken Wharton and the participants of the Journées de Broglie 2023 and the Helgoland 2025 conferences are gratefully acknowledged.

\bibliography{QuantFound}

\end{document}